\documentstyle[12pt]{article}

\newcommand{\s}{\vspace{0.2in}}
\newcommand{\sech} {\rm sech}
\begin{document}
{}~\hfill{UICHEP-TH/93-13}

{}~\hfill{Sept. 8, 1993}

\vspace{.6in}
\centerline {\large\bf Is the Lowest Order Supersymmetric WKB Approximation}\s
\centerline {\large\bf Exact for All Shape Invariant Potentials ?}
\vspace{.5in}

\centerline {D.T. Barclay, Avinash Khare$^1$ and U. Sukhatme}
\vspace{0.3in}
\centerline {Department of Physics, University of Illinois at Chicago,}
\centerline {845 W. Taylor Street, Room 2236 SES,}
\centerline {Chicago, Illinois 60607-7059}
\vspace {1in}

\centerline{\bf Abstract}
\vspace{0.3in}

It has previously been proved that the lowest order supersymmetric WKB
approximation
reproduces the exact bound state spectrum of shape invariant potentials.
We show that this is not true for a new,
recently discovered class of
shape invariant potentials and analyse the reasons underlying
this breakdown of the usual proof.

\vspace{1.5in}
${^1}$ Institute of Physics, Sachivalaya Marg, Bhubaneswar 751005,India.

\newpage

In the past few years, the supersymmetry inspired WKB
approximation (SWKB) [1] has received considerable attention [2].
One reason for this was the proof [3] that the
leading order SWKB quantization
condition reproduces the exact bound state spectra for any shape
invariant potential (SIP) [4]. Subsequently, Adhikari et al.[5] showed by
explicit calculation that the higher order corrections to the energy
eigenvalue spectrum vanish to $O({\hbar}^6)$ for all then-known
shape invariant potentials [6].
For these cases, all the higher-order corrections have since been shown
to vanish [7,8].
The SWKB quantization condition has also been applied to many
non-shape invariant potentials
and it turns out that even though the SWKB formula does better than the
usual WKB approach in most
cases [5,9,10],
it has never been found to be exact for these,
thus suggesting that perhaps shape invariance is not
only sufficient but even necessary for the lowest order SWKB formula
to yield exact energy eigenvalues [10].
Barclay and Maxwell [8] proposed a very simple condition on the
superpotential which ensures
that the lowest order SWKB formula gives exact bound state
spectra with all higher order corrections zero. By analysing this simple
condition, they have suggested that there are no SIPs other than those
tabulated
by Dutt et al.[6] and the one example found by Levai [11]. It should be noted
here that most of these potentials are also contained in the list of Infeld
and Hull [12].

Recently, a large class of new SIPs has been discovered by us [13,14]
disproving the conjecture that
no more examples exist. These new SIPs are reflectionless and possess an
infinite
number of bound states. They can be looked upon as q-deformations of
the symmetric Rosen-Morse potentials corresponding to one or multi-
soliton solutions. Although these new potentials cannot be written in a
closed form using elementary functions, the energy eigenvalue spectrum
is known analytically. It is then  reasonable to enquire if the
lowest order SWKB formula yields the exact bound state spectra for these new
SIPs or not. It is worth emphasizing that the answer to this question
is not obvious. On the one hand, since the lowest order SWKB
formula has been shown
to be exact for all SIPs [3], it is natural to expect that even
for the new SIPs, the lowest order SWKB formula must be exact.
On the other hand,
Barclay and Maxwell [8] have claimed that there are no other potentials
for which the lowest order SWKB formula is exact and the
higher order corrections
are zero. Thus, it is clearly of interest to know if the lowest order
SWKB formula is exact for the new SIPs or not. This is the question that we
address in this letter.

In SUSY quantum mechanics, the partner potentials $V_{\pm}(x,a_0)
=W^2 \pm W'(x,a_0)$ are said to be shape invariant if ($2m = 1$)
$$W^2(x,a_0)+\hbar W'(x,a_0) = W^2(x,a_1)-\hbar W'(x,a_1)+R(a_0),\eqno{(1)}$$
where $W$ is the superpotential, $a_0$ denotes a set of parameters, $a_1=
f(a_0)$ is an arbitrary function of $a_0$ and $R(a_0)$ is independent of $x$.
For all the standard SIPs [6,11] -- for which the lowest order
SWKB formula is exact -- it
turns out that the parameters $a_1$ and $a_0$ are related by translation
($a_1 = a_0 + \alpha \hbar$; $\alpha$ being a constant). On the other hand, for
the new SIPs, $a_1$ and $a_0$ are related by scaling ($a_1 = qa_0$; $0 <q <1$).
Expansion of the superpotential and $R(a_0)$ in powers of $a_0$ gives
$$W(x,a_0) =\sum_{j=0}^{\infty} g_j(x)a_0^j ,~~ R(a_0) =\sum_{j=0}^{\infty}
R_ja_0^j ,~~ a_1 = qa_0. \eqno{(2)}$$
Substitution into the shape invariance condition, eq. (1), and equating
powers of $a_0$ permits a full determination of the quantities $g_j(x)$
for any choice of the constants $R_j$. A particularly simple choice is to
take only one non-zero $R_j$.
It has been shown [13,14] that this special choice gives the
self-similar reflectionless potentials studied by
Shabat [15] and Spiridonov [16]. In that case [13,14]
$$R_n = R_1 \delta_{n1},~~ \beta_1 = {R_1\over (1+q)},\eqno{(3)}$$
$$g_n(x) = \beta_n x^{2n-1} = \bigg ( -{(1-q^n)\over
(2n-1)(1+q^n)}\sum^{n-1}_{j=1} \beta_j \beta_{n-j}\bigg ) x^{2n-1},\eqno{(4)}$$
so that $W(x)$ is an odd function of $x$ and hence SUSY is unbroken. The
exact bound state spectrum for $V_-(x,a_0)$
is now easily calculated and is given by [13,14]
$$E_n = R_1a_0{(1-q^n)\over (1-q)} ,~~  n=0,1, \ldots \eqno{(5)}$$

For the case of unbroken SUSY,
the lowest order quantization condition is
$$ \int\sqrt{E -W^2}~ dx =n \pi \hbar, \quad n=0,1, \ldots \eqno{(6)}$$
By construction this will always give
the exact ground state energy ($E_0 = 0$), but what about the excited
states~?
Since
$W(x)$ is not available in closed form in terms of elementary functions,
the SWKB integral has to be done numerically.
A method for calculating $W(x)$ has been described before [14] and
careful numerical integration shows that for all $q \in (0,1)$ the lowest
order SWKB formula is not exact for the new potentials.
The left hand side of
equation (6) can be thought of as defining a function
$n(E) \equiv (\pi \hbar)^{-1} \int \sqrt{E-W^2} dx$ which will take
on the integer value $n$ for $E=E_n$, if this lowest order quantization
condition
is exact.
In Table 1 we display the results of doing the integral using the
exact results for $E=E_1$, i.e. calculating $n(E_1)$.
For $q$ close to one the discrepancy is small (though still significantly
greater than
the numerical uncertainties), but it diverges (roughly as $\ln q$)
as $q \rightarrow 0$. At exactly $q=0$, the Rosen-Morse potential,
the condition is known to be exact, a discontinuity that is presumably a
reflection of the way in which the infinitely many excited states for $q>0$
collapse to a single one when $q=0$.

The limit $q \rightarrow 1$ is much smoother and corresponds to the harmonic
oscillator, for which the SWKB formula is also known to be exact.
In fact, it appears that one can perturb about this limit for small values of
$\epsilon \equiv 1-q$.
The shape invariance condition (1) can be expanded to find $W$ as a power
series in $\epsilon$ and this used to evaluate the integral. One finds that
$$ n(E_1) = 1 + {83 \over 648} \epsilon^2 + O(\epsilon^3), \eqno{(7)}$$
which accords well with the numerical results close to $q=1$.
All the above calculations are done for the simplest
case where only $R_1$ is taken non-zero. It is easily checked that
the lowest order SWKB formula does not give correct eigenvalues
for the more general and complicated cases of new SIPs corresponding to
several non-zero $R_j$ either. This is of course intuitively expected.

What is the special feature of these new potentials that interferes with the
proof that
condition (6) is exact for shape invariant potentials~?
To understand this, let
us take a fresh look at the derivation of the lowest order SWKB quantization
condition, paying particular attention to the role of $\hbar$.
The lowest order WKB quantization condition for the
potential $V_-(x,a_0$) is ($2m=1$)
$$  \int \sqrt{E_n -V_-(x,a_0)} \quad dx = (n+1/2) \pi \hbar. \eqno{(8)}$$
In terms of the superpotential $W$, this reads
$$ \int\sqrt{E_n -W^2(x,a_0)+\hbar W'(x,a_0)} \quad dx = (n+1/2) \pi \hbar.
\eqno{(9)}$$
Now comes the crucial step in the derivation. One argues that formally
$W^2$ is of $O(\hbar^0)$ while $\hbar W'$ is of $O(\hbar)$ and
hence one can expand the integrand on the left hand side
in powers of $\hbar$ to get the lowest order SWKB quantization condition (6).
This assumption is justified for all the standard SIPs [6,11]
since $W^2$ is indeed of $O(\hbar^0)$ while $\hbar W'$ is indeed of
$O(\hbar)$. One might object to this procedure since the resulting potential
$V_-$  is then $\hbar$-dependent. However, in every
one of those cases this $\hbar$-dependence can be absorbed into
some dimensionful parameters in the problem.
For example, consider
$$W = A \tanh x  \eqno{(10)}$$
so that
$$V_-(x) = A^2 - A(A+\hbar) \sech^2x.   \eqno{(11)}$$
Taking $A$ such that $A(A+\hbar)$ is independent of $\hbar$ gives the desired
$\hbar$-independent potential (the additive constant is irrelevant and
so can contain $\hbar$).
Such a move may appear to be of limited value since
one cannot apply SWKB directly to a superpotential $W$ which is
now $\hbar$-dependent.
However, because $A$ is a free parameter, one can continue the SWKB results
obtained
for $A$ (and hence $W$) independent of $\hbar$ over to this superpotential and
so obtain an SWKB approximation for a $\hbar$-independent potential.

What about the new potentials~?
In the simplest of these cases, the only free parameter in the problem (apart
from $q$) is the combination $R_1 a_0$, on which $W$ depends as $W(x, R_1 a_0)
=\sqrt{R_1 a_0} F(\sqrt{R_1 a_0} x/\hbar)$.
Incorporating different dependences on $\hbar$ in $R_1 a_0$ will give
different ones in $W$, $V_-$ and $E_n$, but $F$ is a sufficiently
complicated function that there is no way of eliminating $\hbar$ from $W^2$.
This is a direct consequence of the  scaling reparameterisation $a_1=q a_0$
not involving $\hbar$: if $W^2(x,a_0)$ were independent of $\hbar$, so would
$W^2(x,a_1)$ be and in taking the lowest order of (1) one would get
$W^2(x,a_0)=W^2(x,a_1)$, which corresponds to the harmonic oscillator.
A reparameterisation of the form $a_1= a_0 + \alpha \hbar$ clearly
avoids this.
Furthermore, with $a_1=q a_0$, $W^2$ and $\hbar W'$ are now of a similar
order in $\hbar$.
The basic distinction between them involved in deriving equation (6) is thus
no longer valid and we are prevented from deriving the SWKB condition for
these new potentials.
Similar difficulties will arise for potentials with more $R_j$ non-zero
and for the other $\hbar$-independent
reparameterisations considered in [14].

It might be thought that an alternative derivation of the SWKB series for these
potentials could be constructed by considering $W=\sqrt{R_1 a_0}
F(\sqrt{R_1 a_0} x/H)$, where $H$ is simply a free constant unrelated to
$\hbar$.
The series can be safely derived for all $H$ since $W$ is $\hbar$-independent
and the result continued to $H=\hbar$ at the end.
Unfortunately, when this is done it becomes clear that the structure of all
the terms is such that the entire series is then all of $O(\hbar)$, so that
one cannot usefully separate out a lowest order condition corresponding to
(6) from a set of higher order corrections.
It is then little surprise that an arbitrarily selected piece (6) of the full
condition does not give the exact spectrum.

Since it appears to have been overlooked in the extensive literature
(see e.g. [17]) on symmetric, reflectionless potentials containing infinitely
many bound states, we point out that the inextricable $\hbar$-dependence
of these new shape invariant potentials is actually a general consequence
of them possessing these properties.
Reflectionlessness implies that a potential is finitely deep, and the infinite
spectrum means that the top of it is given by the limit of $E_n$ as
$n \rightarrow \infty$.
However, $E_n \rightarrow 0$ as $\hbar \rightarrow 0$, so the potential
must also distort in the semi-classical limit; a little further reflection
shows that it always tends towards a free-particle potential.
A similar conclusion can be reached for the $N$-soliton reflectionless
potential (which contains $N$ bound states) by
explicitly restoring $\hbar \neq 1$
in the inverse scattering formalism for these potentials [17].

We have thus seen that the SWKB quantization condition (6) is
not the correct lowest order formula in the case of the new SIPs and hence
it is not really surprising that eq. (6) does not give the exact eigenvalues
for
these potentials.
In other words, it remains true that the lowest order SWKB quantization
condition is
exact for SIPs (if the SUSY is unbroken), but only in those cases
for which the formula is applicable in the first place.
It is thus still the case that the SIPs given in refs.[6,11]
are the only known ones for which the lowest order SWKB formula is
exact and the higher order corrections are all zero.

It is a pleasure to thank R. Dutt, A. Gangopadhyaya, C.J. Maxwell and
A. Pagnamenta for helpful discussions. This work was supported in part by the
U. S. Department of Energy under grant DE-FGO2-84ER40173.

\vfill
\eject

\vspace{2cm}

\begin{center}
\begin{tabular}{|c|c|}\hline
$\quad \qquad q \qquad \quad$ &$ \quad \qquad n(E_1) \quad \qquad $ \\
\hline
0.01 & 1.238 \\
0.1 & 1.101 \\
0.2 & 1.063 \\
0.3 & 1.042 \\
0.4 & 1.029 \\
0.5 & 1.019 \\
0.6 & 1.012 \\
0.7 & 1.007 \\
0.8 & 1.003 \\
0.9 & 1.001 \\
\hline
\end{tabular}
\end{center}
\vspace{0.5cm}
{\bf Table 1:} Results of evaluating the lowest order SWKB integral
using the exact value $E_1=R_1 a_0$ for different values of $q$. An
exact quantisation condition would give $n=1$ in every case. Note that
the $R_1 a_0$ dependence cancels in the integral, so these results do
not depend on this free parameter.

\end{document}